 \documentclass[aps,pra,preprint,groupedaddress,showpacs]{revtex4}
 \usepackage{graphicx} 
 \usepackage{dcolumn}  
 \usepackage{bm}       

\begin{document}
\preprint{} 

\title
{Vacuum polarization calculations for hydrogenlike and \\ alkalilike ions}

\author{J. Sapirstein}
\email[]{jsapirst@nd.edu}
\affiliation
{Department of Physics, University of Notre Dame, Notre Dame, IN 46556}

\author{K. T. Cheng}
\email[]{ktcheng@llnl.gov}
\affiliation
{University of California, Lawrence Livermore National Laboratory,
Livermore, CA 94550}

\date{\today}

\begin{abstract}
Complete vacuum polarization calculations incorporating finite nuclear size
are presented for hydrogenic ions with  principal quantum numbers $n=1-5$.
Lithiumlike, sodiumlike, and copperlike ions are also treated starting
with Kohn-Sham potentials, and including first-order screening corrections.
In both cases dominant Uehling terms are calculated with high accuracy,
and smaller Wichmann-Kroll terms are obtained using numerical electron
Green's functions.
\end{abstract}


\maketitle

\section{Introduction}

The most important radiative correction to the spectra of highly charged
ions is the Lamb shift, which at the one-loop level consists of two
contributions, the self-energy and vacuum polarization. The effect
of the latter is dominated by the Uehling term \cite{Uehling},
which was introduced even before the development of the modern form
of Quantum Electrodynamics (QED). However, while relatively small,
extra effects known as Wichmann-Kroll terms \cite{WK} need to be
considered when theory is confronted by experimental data on
highly-charged ions of ever-increasing accuracy. An important advance
in the field was made by Gyulassy \cite{Gyulassy}, who pointed out that
by arranging the calculation so that electron propagator terms with
positive and negative values of the angular momentum quantum number
$\kappa$ are taken together, a single subtraction
of the Uehling term leads to a finite expression for
the Wichmann-Kroll term free of the spurious constants that can arise
when formally infinite integrals are manipulated. He presented values
only for $|\kappa| =1$, but Soff and Mohr \cite{Soff-Mohr} then treated
higher $\kappa$ values and showed that the partial wave series
converges very rapidly.

The results of Soff and Mohr covered $n=1$ and $n=2$ states, and included
finite nuclear size through a shell model, which allowed the use of analytic
Green's functions. However, many-electron ions are best described with
non-Coulomb potentials that incorporate electron screening, and in this
case one is forced to use numerical Green's functions. We have used these
functions extensively in treating the self-energy \cite{CJS}, and one purpose
of this paper is to extend their use to the evaluation of the Wichmann-Kroll
terms in a variety of atomic systems.

Other work in the field includes a basis set based approach by
Persson {\it et al.}~\cite{Persson} that can also be applied to non-Coulomb
potentials. Ref.~\cite{Persson} in particular calculates the Wichmann-Kroll
effect for lithiumlike uranium, which will also be treated here. We note
the work of Beier {\it et al.}~\cite{Beier} for hydrogenic ions,
which is of particularly high accuracy. Finally, a calculation of
vacuum polarization effects on lithiumlike ions, very similar in its
treatment of screening to what will be presented here, can be found in
Ref.~\cite{Shabaev}.

In the next section, we briefly describe our treatment of the dominant
Uehling term. In the following section, formulas for the Wichmann-Kroll
term are given, along with a discussion of numerical issues. Screening
terms are then defined, and results are given in a set of tables.
A brief discussion of the accuracy of our results for lithiumlike,
sodiumlike, and copperlike uranium completes the paper.

\section{Formalism}

We work in Furry representation QED \cite{Furry}, with a lowest-order
Hamiltonian given by
\begin{equation}
  H_0 = \int\! d^3 x\, \psi^{\dagger}(x)
  \bigg[\! -i \vec{\alpha} \cdot \vec{\nabla} +
  \beta m - {Z_{\rm eff}(r) \alpha \over r}\,\bigg] \psi(x) \, ,
\end{equation}
where $r=|\vec x|$. Natural units in which $\hbar = c = 1$ are used here.
In the usual Furry representation, $Z_{\rm eff}(r) = Z_{\rm nuc}(r)$
describes the Coulomb field of the nucleus, which will be modeled here
with a Fermi distribution. However, when describing many-electron atoms
and ions in a QED framework, it is often convenient to incorporate some
screening through the use of non-Coulomb potentials. Here we work with
three effective charges, designed to incorporate screening effects
in lithiumlike, sodiumlike, and copperlike ions. They represent
Kohn-Sham potentials, defined through
\begin{equation}
  Z_{\rm eff}(r) = Z_{\rm nuc}(r) - r \!\int\! dr' {1 \over r_>} \rho_t(r')
  + {2 \over 3}\, \bigg[ {81~ \over 32 \pi^2} \, r \rho_t(r) \bigg]^{1/3},
\label{K-S}
\end{equation}
where
\begin{equation}
  \rho_t(r) = g^2_v(r) + f^2_v(r) +
  \sum_a \, (2j_a + 1) \left[ g_a^2(r) + f_a^2(r) \right]
\end{equation}
is the electronic charge density such that $\int\!\rho_t(r) dr = N$,
the total number of electrons. Here $g(r)$ and $f(r)$ are the upper and
lower components of Dirac wave functions determined self-consistently,
and the indices $v$ and $a$ refer to valence and core electrons,
respectively. For lithiumlike ions, $v=2s$ and $a=1s_{1/2}$ state.
For sodiumlike ions, $v=3s$ and $a$ ranges over the $2s_{1/2}$, $2p_{1/2}$,
and $2p_{3/2}$ states also. Finally for copperlike ions, $v=4s$, and $a$
includes the additional $3s_{1/2}$, $3p_{1/2}$, $3p_{3/2}$, $3d_{3/2}$,
and $3d_{5/2}$ core states.

The basic expression for vacuum polarization is given by
\begin{equation}
  E^{\rm VP} = - i \alpha \int {dk_0 \over 2 \pi} \int\! d^3 x d^3 y \,
  \bar{\psi}_v(\vec x) \gamma_0 \psi_v(\vec x) { 1 \over | \vec x - \vec y|}
  {\rm Tr} \left[ \gamma_0 S_F(\vec y, \vec y; k_0) \right],
\label{basic}
\end{equation}
where the bound electron propagator $S_F$ satisfies the equation
\begin{equation}
  \bigg[ E + i \vec{\alpha} \cdot \vec{\nabla} - \beta m +
  {Z_{\rm eff}(x) \alpha \over x} \, \bigg] S_F(\vec x, \vec y; E) =
  \delta^3(\vec x - \vec y) \gamma_0 \, .
\end{equation}
We will also need the free electron propagator $S_0(\vec x, \vec y; E)$,
which satisfies
\begin{equation}
  \left[ E  + i \vec \alpha \cdot \vec \nabla -\beta m \right]
  S_0(\vec x, \vec y; E) = \delta^3(\vec x - \vec y) \gamma_0 \, ,
\end{equation}
and is solved by
\begin{equation}
  S_0(\vec x, \vec y; E) = \int {d^3 p \over (2\pi)^3} \,
  {e^{i \vec p \cdot (\vec x - \vec y)} \over
  E \gamma_0 - \vec p \cdot \vec \gamma - m} \, .
\label{S0def}
\end{equation}
The bound propagator $S_F$ can be expanded in terms of the free propagator
$S_0$ through
\begin{equation}
  S_F(\vec x, \vec z; E) = S_0(\vec x, \vec z; E) -
  \int\! d^3 y \, S_0(\vec x, \vec y; E) {Z_{\rm eff}(y) \alpha \over y}
  \gamma_0 S_F(\vec y, \vec z;E) .
\label{basics}
\end{equation}

\subsection{Uehling Term}

When the above expansion of the propagator is used in Eq.~(\ref{basic}),
the first term on the right hand side makes no contribution because of
Furry's theorem \cite{Furryt}, which states that there is an exact
cancellation between loop diagrams with an odd number of external
photons when the momentum in the loop is reversed, which follows from
a charge conjugation argument. When we set up the numerical calculation
of the Wichmann-Kroll terms below using a partial wave expansion of the
electron propagator, we will again discuss the vanishing of this term.
The Uehling approximation to vacuum polarization comes from replacing
the last bound propagator $S_F$ in Eq.~(\ref{basics}) with a free
propagator $S_0$, which leads to the
ultraviolet divergent energy shift
\begin{eqnarray}
  E^{\rm VP}_{\rm Ueh} &=& {i \alpha \over 2 \pi}
  \int\! dk_0 \int\! d^3 x\, d^3 y\, d^3 z\,
  \bar{\psi}_v(\vec x) \gamma_0 \psi_v(\vec x)
  {1 \over |\vec x - \vec y|}
 \nonumber \\ && \times
  {\rm Tr} \bigg[ \gamma_0 S_0(\vec y, \vec z; k_0)
  { Z_{\rm eff}(z) \alpha \over z}
  \gamma_0 S_0(\vec z, \vec y; k_0) \bigg] .
\label{startuel}
\end{eqnarray}
It is useful at this point to define the vacuum-polarization tensor in
$n=4-\epsilon$ dimensions,
\begin{equation}
  \Pi_{\mu \nu}(q) = -i e^2 \int\! {d^n k \over (2\pi)^n} \,
  {\rm Tr} \bigg[ \gamma_{\mu}
  { 1 \over \not\!q + \not\!k -m} \gamma_{\nu}
  { 1 \over \not\!k - m} \bigg],
\end{equation}
which is ultraviolet finite for positive $\epsilon$.
An advantage of this so-called dimensional regularization is that an
automatically gauge-invariant form results after combining denominators
with a Feynman parameter $u$ and carrying out the $d^n k$ integration,
\begin{equation}
  \Pi_{\mu \nu}(q) = \Big( q^2 g_{\mu \nu} - q_{\mu} q_{\nu} \Big)
  \bigg\{ Z_3^{(2)} - {2 \alpha \over \pi} \int_0^1\! du \, u(1-u)
  \, {\rm ln}\Big[1 -u (1-u) q^2/m^2 \Big] \bigg\}.
\end{equation}
The constant $Z_3^{(2)}$, which diverges as $1/\epsilon$, is removed
by the renormalization procedure, leaving a finite expression. In the
case of vacuum polarization the relevant 4-vector $q$ has a vanishing
time component, and $E^{\rm VP}_{\rm Ueh}$ depends only the $\mu = \nu = 0$
component,
\begin{equation}
  \Pi_{00}(q) = {2 \alpha \over \pi} \, \vec q\,^2 \!\int_0^1\! du \,
  u(1-u) \, {\rm ln} \Big[ 1 + u (1-u) \vec q\,^2/m^2 \Big].
\end{equation}
If we now use the representation of the free propagator given in
Eq.~(\ref{S0def}) for the Uehling energy $E^{\rm VP}_{\rm Ueh}$ shown
in Eq.~(\ref{startuel}), an expression involving the vacuum polarization
tensor results, allowing us to write
\begin{eqnarray}
  E^{\rm VP}_{\rm Ueh} &=& -{\alpha \over 2 \pi^2}
  \int\! d^3 x\, d^3 y\, d^3 z\, \psi^{\dagger}_v(\vec x) \psi_v(\vec x) \,
  {1 \over |\vec x - \vec y \,|} \, {Z_{\rm eff}(z) \alpha \over z}
 \nonumber \\ && \times
  \int\! {d^3 q \over (2 \pi)^3} \, e^{i \vec q \cdot (\vec y - \vec z)}
  \vec q\,^2 \!\int_0^1\! du \, u(1-u) \,
  {\rm ln} \Big[ 1 + u (1-u) \vec q\,^2 / m^2 \Big]\,.
\end{eqnarray}
Carrying out the $d^3 y$ integration gives
\begin{eqnarray}
  E^{\rm VP}_{\rm Ueh} &=& - {2 \alpha \over \pi}
  \int\! d^3 x\, d^3 z\, \psi^{\dagger}_v(\vec x) \psi_v(\vec x)
  {Z_{\rm eff}(z) \alpha \over z}
 \nonumber \\ && \times
  \int\! {d^3 q \over (2 \pi)^3} \, e^{i \vec q \cdot (\vec x - \vec z)}
  \!\int_0^1\! du \, u(1-u) \,
  {\rm ln} \Big[ 1 + u (1-u) \vec q\,^2/m^2 \Big]\,.
\end{eqnarray}
A partial integration with respect to $u$ gives a factor ${\vec q}\,^2$
that can be written as $-\nabla_z^2$ operating on the exponential. Two
additional partial integrations with respect to
$\nabla_z$ then leads to
\begin{eqnarray}
  E^{\rm VP}_{\rm Ueh} &=& -{\alpha \over \pi}
  \int\! d^3 x\, d^3 z\, \psi^{\dagger}_v(\vec x) \psi_v(\vec x)
  \nabla_z^2 \bigg[{Z_{\rm eff}(z) \alpha \over z}\bigg]
 \nonumber \\ && \times
  \int\! {d^3 q \over (2 \pi)^3} \, e^{i \vec q \cdot (\vec x - \vec z)}
  \!\int_0^1\! du \, u^2 (1-2u) \left[ 1 - {2u \over 3} \right]
  {1 \over m^2 + u (1-u) \vec q\,^2}\,,
\end{eqnarray}
which results in
\begin{equation}
  E^{\rm VP}_{\rm Ueh} = -{\alpha \over 4\pi^2}
  \int\! d^3 x\, \psi^{\dagger}_v(\vec x) \psi_v(\vec x)
  \int\! d^3 y\, \nabla_y^2 \bigg[ {Z_{\rm eff}(y) \alpha \over y} \bigg]
  \int_0^1\! du \, {u (1-2u) (1-{2u \over 3}) \over 1-u} \,
  {e^{-m|\vec x - \vec y| \over \sqrt{u(1-u)}} \over |\vec x - \vec y|}\,.
\end{equation}
This is equivalent to an alternative form sometimes presented for the
Uehling term,
\begin{equation}
  E^{\rm VP}_{\rm Ueh} = {\alpha \over 4\pi^2}
  \int\! d^3 x\, \psi^{\dagger}_v(\vec x) \psi_v(\vec x)
  \int\! d^3 y\, \nabla_y^2 \bigg[ {Z_{\rm eff}(y) \alpha \over y} \bigg]
  \int_0^1\! dv\, {v^2 (1-{v^2 / 3}) \over 1-v^2} \,
  {e^{-2m|\vec x - \vec y| \over \sqrt{(1-v^2)}} \over |\vec x - \vec y|}\,,
\end{equation}
as can be seen by setting $u=(1-v)/2$ in the region $u=0$ to 1/2 and
$u=(1+v)/2$ in the region $u=1/2$ to 1. When the effective charge is
spherically symmetric, as will be assumed for all cases considered here,
the alternative form can be rewritten, after carrying out the angle
integration in the $d^3 y$ integration and making the final change of
variable $v = \sqrt{1-1/t^2}$,
\begin{eqnarray}
  E^{\rm VP}_{\rm Ueh} &=& {\alpha^2 \over 6 \pi m}
  \int\! d^3 x\, \psi^{\dagger}_v(\vec x) \psi_v(\vec x) \, {1 \over x}
  \int_0^{\infty}\! d y\, Z''_{\rm eff}(y)
 \nonumber \\ && \times
  \int_1^{\infty}\! dt \, \sqrt{t^2-1} \,
  \bigg( {1 \over t^2} + {1 \over 2 t^4} \bigg)
  \Big[ e^{-2m|x - y| t} - e^{-2m(x+y) t} \Big].
\end{eqnarray}
The $t$ integration has been treated by Fullerton and Rinker \cite{Full-Rin},
who give numerical fitting formulas that allow fast and accurate
evaluation of its value, after which a single numerical $y$ integration
gives the Uehling potential $U_V(x)$ as defined by
\begin{equation}
  E^{\rm VP}_{\rm Ueh}
  \equiv \int\! d^3 x\, \psi^{\dagger}_v(\vec x) U_V(x) \psi_v(\vec x)
  \equiv (U_V)_{vv} \, ,
\end{equation}
such that the Uehling energy is given by the expectation value of
this potential. Before presenting results for the contribution of
the Uehling terms to vacuum polarization, we turn to a description
of the calculation of Wichmann-Kroll terms.

\subsection{Wichmann-Kroll terms}

With the Uehling term accounted for, the remaining part of the vacuum
polarization is given by the Wichmann-Kroll term,
\begin{eqnarray}
  E^{\rm VP}_{\rm WK} &=& -{i \alpha \over 2 \pi}
  \int\! dk_0 \int\! d^3 x\, d^3 y\,
  \psi^{\dagger}_v(\vec x) \psi_v(\vec x) \,
  {1 \over |\vec x - \vec y|} \,
  {\rm Tr} \bigg\{ \gamma_0 \Big[ S_F(\vec y, \vec y; k_0)
 \nonumber \\ &&
  - \: S_0(\vec y, \vec y; k_0) + \!\int\! d^3 z\,
  S_0(\vec y, \vec z; k_0) \, {Z_{\rm eff}(z) \alpha \over z} \,
  \gamma_0 S_0(\vec z, \vec y; k_0) \Big] \bigg\} \, .
\end{eqnarray}
While this combination is ultraviolet finite by power counting, a spurious,
non-gauge invariant finite term can show up if sufficient care is not taken.
As was shown by Gyulassy \cite{Gyulassy}, this problem can be avoided when
the propagators are represented with a partial wave expansion by an
appropriate grouping of that expansion. The calculation is carried out
as follows. We first make a Wick rotation $k_0 \rightarrow i \omega$.
Unlike the case of the self-energy, no poles are passed in this process.
The next step is to replace the electron propagators with the partial
wave expansion
\begin{equation}
  S_F(\vec x, \vec y; E) = \sum_{\kappa,\mu} \, \left[
  \theta(x-y) w^E_{\kappa \mu}(\vec x) \bar{u}^E_{\kappa \mu}(\vec y) +
  \theta(y-x) u^E_{\kappa \mu}(\vec x) \bar{w}^E_{\kappa \mu}(\vec y)
  \right] .
\end{equation}
Here, $\theta(x)$ is the step function such that $\theta(x) = 1$ if
$x > 0$ and $\theta(x) = 0$ if $x < 0$, and
$w^E_{\kappa \mu}$ and $u^E_{\kappa \mu}$ are the solutions of
the Dirac equation regular at infinity and the origin, respectively.
They are represented as
\begin{equation}
  u^E_{\kappa \mu}(\vec r) = {1 \over r}
  \left(
    \begin{array}{rl}
      i\,g^0_{\kappa}(E,r) & \chi_{ \kappa \mu} (\Omega) \\
         f^0_{\kappa}(E,r) & \chi_{-\kappa \mu} (\Omega)
    \end{array}
  \right),
\end{equation}
and
\begin{equation}
  w^E_{\kappa \mu}(\vec r) = {1 \over r}
  \left(
    \begin{array}{rl}
      i\,g^{\infty}_{\kappa}(E,r) & \chi_{ \kappa \mu} (\Omega) \\
         f^{\infty}_{\kappa}(E,r) & \chi_{-\kappa \mu} (\Omega)
    \end{array}
  \right).
\end{equation}
The unsubtracted vacuum polarization is, in terms of this expansion
of the electron propagator,
\begin{eqnarray}
  E^{\rm VP} &=& {\alpha \over 2 \pi} \int_{-\infty}^{\infty} d\omega
  \int {d^3 x\, d^3 y \over |\vec x - \vec y|} \,
  \psi_v^{\dagger}(\vec x) \psi_v(\vec x) \, {1 \over y^2}
  \sum_{\kappa \mu} {\rm Tr} \left[
  \gamma_0 u_{\kappa \mu}^{i\omega}(\vec y)
  \bar{w}_{\kappa \mu}^{i\omega}(\vec y) \right]
 \nonumber \\ &=&
  {\alpha \over 2 \pi} \int_{-\infty}^{\infty} d\omega
  \int {d^3 x\, d^3 y \over |\vec x - \vec y|} \,
  \psi_v^{\dagger}(\vec x) \psi_v(\vec x) \, {1 \over y^2}
  \sum_{\kappa \mu} \bigg\{
  g^{\infty}_{\kappa}(i\omega,y) g^0_{\kappa}(i\omega,y)
  {\rm Tr} \Big[ \chi_{\kappa \mu}(\Omega_y)
       \chi^{\dagger}_{\kappa \mu}(\Omega_y) \Big]
 \nonumber \\ &&
  + \, f^{\infty}_{\kappa}(i\omega,y) f^0_{\kappa}(i\omega,y)
  {\rm Tr} \Big[ \chi_{-\kappa \mu}(\Omega_y)
       \chi^{\dagger}_{-\kappa \mu}(\Omega_y) \Big] \bigg\}.
\end{eqnarray}
If we then use the identity
\begin{equation}
  {\rm Tr} \Big[ \sum_{\mu} \chi_{\kappa \mu}(\Omega)
  \chi^{\dagger}_{\kappa \mu}(\Omega) \Big] = {|\kappa| \over 2 \pi}
\end{equation}
we have the compact expression
\begin{equation}
  E^{\rm VP} = {\alpha \over 4 \pi^2} \sum_{\kappa} |\kappa|
  \!\int_{-\infty}^{\infty}\! d\omega
  \!\int\! {d^3 x\, d^3 y \over |\vec x - \vec y|} \,
  \psi_v^{\dagger}(\vec x) \psi_v(\vec x) {1 \over y^2} \,
  \Big[ g^{\infty}_{\kappa}(i\omega,y) g^0_{\kappa}(i\omega,y) +
        f^{\infty}_{\kappa}(i\omega,y) f^0_{\kappa}(i\omega,y) \Big].
\end{equation}

Before continuing to the subtraction of the Uehling term, we discuss
the subtraction of the free propagator term, which vanishes by
Furry's theorem \cite{Furryt}. In terms of the partial wave expansion
of the free propagator term, which is given by the same expression as
above but with the bound radial Green's functions $g^{0,\infty}_\kappa$
and $f^{0,\infty}_\kappa$ replaced by the corresponding free-electron
functions, the manifestation of this cancellation is particularly simple,
as each positive value of $\kappa$ gives a contribution canceled by the
corresponding negative value. This was shown analytically by Gyulassy
\cite{Gyulassy}, but here, since we treat both the bound and free
propagators with the same numerical methods, we use it as a check on
the accuracy of our Green's functions. In Table \ref{tab1}, we present
partial wave results of the free-propagator term $E^{\rm VP}(S_0)$
for low values of $\kappa$ for the finite nuclear size $1s$ ground
state of hydrogenlike mercury ($Z=80$). The cancellation of this term
between the positive and negative $\kappa$ partial waves is obvious.
Only six digits past the decimal point are shown, but the actual
cancellation is even finer, which is one of the numerical tests used
in this work. Also shown in the same table are values of the bound
propagator term $E^{\rm VP}$ which can be seen to scale roughly with
$|\kappa|$, which is how a quadratic ultraviolet divergence is
manifested in a partial wave expansion.

We next turn to the unrenormalized Uehling term, which must be evaluated
with high accuracy, as it cancels out many digits of the unrenormalized
vacuum polarization. While it can be formed analytically, we have
found that numerical methods actually work somewhat better, and
continue to use them. A short calculation gives the Uehling term as
\begin{eqnarray}
  E^{\rm VP}_{\rm Ueh} &=& -{2 \alpha \over \pi} \,
  Re \int_0^{\infty}\! d\omega \!\int_0^{\infty}\! dr \,
  \Big[g_v(r)^2 + f_v(r)^2 \Big] \!\int_0^{\infty}\! dr' {1 \over r_>}
  \int_0^{\infty}\! dx \, { Z_{\rm eff}(x) \alpha \over x}
  \sum_{\kappa} |\kappa|
 \nonumber \\ && \times
  \bigg\{ \theta(r'-x)
   \Big[ g^0_\kappa(i\omega,x)^2  + f^0_\kappa(i\omega,x)^2 \Big]
   \Big[ g^\infty_\kappa(i\omega,r')^2 + f^\infty_\kappa(i\omega,r')^2 \Big]
 \nonumber \\ &&
  \,\,\,+\, \theta(x-r')
   \Big[ g^\infty_\kappa(i\omega,x)^2 + f^\infty_\kappa(i\omega,x)^2 \Big]
   \Big[ g^0_\kappa(i\omega,r')^2 + f^0_\kappa(i\omega,r')^2 \Big] \bigg\}.
\end{eqnarray}
To carry out the evaluation of the partial wave expansion form of the
Uehling term with sufficient accuracy requires a great deal of care.
Matters are facilitated by using an extremely fine radial grid of up
to 50,000 points. This is particularly helpful in controlling the
accuracy of the $\omega$ integration where the virtual infinity
of the calculation can exceed $1\!\times\!10^6$ mc$^2$
and numerical instabilities at the highest $\omega$ values can lead
to unphysical oscillations in the renormalized effective charge density
at very small and very large radial points. These same problems play
a much smaller role in self-energy calculations, where the bound state
wave function provides suppression at these regions: here, however,
the wave function does not provide any suppression.

The second test of the numerics of our approach has to do with the
integral over the effective charge density $\rho(r)$, defined through
\begin{equation}
  E^{\rm VP}_{\rm WK} = \int_0^{\infty}\! dr \rho(r) \phi(r),
\end{equation}
where
\begin{equation}
  \phi(r) = \int_0^{\infty}\! dr' {1 \over r_>}
  \Big[ g_v(r')^2 + f_v(r')^2 \Big] .
\end{equation}
Without the screening potential $\phi(r)$ from the bound electron $v$,
the effective charge density $\rho(r)$ should integrate to zero,
as there should not be any change in the net charge. Contributions
to this integral from the unrenormalized vacuum polarization and
Uehling terms cancel to a level that is typically of order $10^{-8}$
or better. If this cancellation is not this precise, we have found
that our answers become unstable and go into disagreement with
previous calculations.

Partial wave results of the unrenormalized Uehling term for the $1s$
state of Hg$^{79+}$ are shown also in Table \ref{tab1}. It can be seen
that they are independent of the sign of $\kappa$ and show the same
rate of increase with $|\kappa|$ as the bound propagator term
$E^{\rm VP}$. The quadratic ultraviolet divergence in these term
cancels, however, and their sum gives the finite Wichmann-Kroll
term which can be seen to converge very rapidly with $|\kappa|$.

We now present results for the hydrogenic vacuum polarization
including finite nuclear size for states of principal quantum
number $n=1-5$, excluding $d$ and higher angular momentum states
which have very small contributions. We model the nuclear charge
distribution with the Fermi distributions described in
Ref.~\cite{Johnson-Soff}, with the exception of the cases $Z=90$
and $Z=92$, where we use $c=7.0589$ and $c=7.13753$, respectively,
which were derived from Ref.~\cite{Zumbro90,Zumbro92}. It is
convenient to pull out the overall $Z$ and $n$ behavior by working
in terms of the function $F_n(Z\alpha)$ defined through
\begin{equation}
  E^{\rm VP} = {\alpha \over \pi} \, {(Z \alpha)^4 \over n^3} \,
  F_n(Z \alpha) \, mc^2 \, .
\end{equation}
Results for $ns$, $np_{1/2}$ and $np_{3/2}$ states are given in
Tables \ref{tab2} -- \ref{tab4}.
Where comparison is possible, we find our results to be in good
agreement with previous calculations cited in the introduction.

\subsection{Screening Corrections}

We now turn to the evaluation of vacuum polarization in
many-electron ions with an alkalilike electronic configuration.
Formulas for the Uehling and Wichmann-Kroll terms given above
are valid for any effective charge $Z_{\rm eff}(r)$, so while
it would be possible to always start with a Coulomb potential
and calculate screening effects starting from that point,
we choose here to incorporate the dominant effect of screening
for the lithium, sodium, and copper isoelectronic sequences
by using the Kohn-Sham potentials defined in Eq.~(\ref{K-S}).
While this is adequate for the Wichmann-Kroll terms, we account
for screening more fully for the Uehling terms in this section.
The theory for lithiumlike ions was set out in some detail in
\cite{LiBi}, so we simply briefly generalize it here to the
sodiumlike and copperlike sequences.

\begin{figure}[h]
\centerline{\includegraphics[scale=0.8]{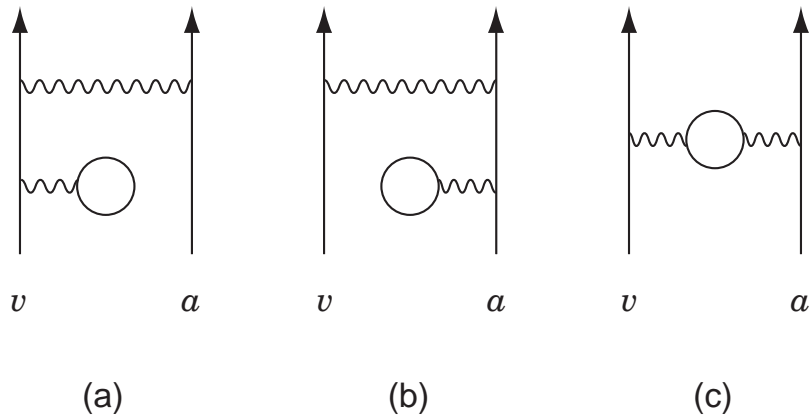}}
\caption{\label{fig1}
Feynman diagrams for the screening correction to vacuum polarization.
}
\end{figure}

There are four sources of screening for the Uehling term:
valence perturbed orbital terms, core perturbed orbital terms,
insertions of the Uehling term in one-photon exchange, and
derivative terms. Representative Feynman diagrams for the
first three sources are shown in Figs.~\ref{fig1}a -- \ref{fig1}c,
though the derivative terms do not have a standard diagrammatic
representation. Beginning with the insertion of the Uehling term
in one-photon exchange, we note that the first-order energy of
an alkalilike ion is given by
\begin{equation}
  E^{(1)} = \sum_a \big[ g_{vava}(0) - g_{vaav}(\delta E_{va})
  \big] - U_{vv} \,,
\end{equation}
where $\delta E_{va} = \epsilon_v - \epsilon_a$,
$U(r) = \big[Z_{\rm nuc}(r) - Z_{\rm eff}(r)\big]\alpha / r$
is the counter potential, and
\begin{equation}
  g_{ijkl}(E) = \alpha \!\int\! d^3 x\, d^3 y\,
  {e^{i E |\vec x - \vec y|} \over |\vec x - \vec y|} \,
  \bar{\psi}_i(\vec x) \gamma^{\mu} \psi_k(\vec x) \,
  \bar{\psi}_j(\vec y) \gamma_{\mu} \psi_l(\vec y) .
\end{equation}
The insertion effect shown in Fig.~\ref{fig1}c can be accounted
for by replacing the $g$ factors in the above, which involve the
exchange of a massless photon, with a weighted integration over
$g$ factors involving the exchange of a massive photon.
The explicit form is
\begin{equation}
  E^{\rm VP}_{\rm Ex} = {\alpha \over \pi} \int_0^1\! du\,
  {u^2 (1 -u^2/3) \over 1-u^2} \, \sum_a \bigg[ g_{vava}(i\xi)
  - g_{vaav}\Big(i\sqrt{\xi^2 - \delta E_{va}^2}\,\Big) \bigg],
\end{equation}
where $\xi = \sqrt{4m^2/(1-u^2)}$. We compared our results
for this term with results for lithiumlike uranium given in
Ref.~\cite{Shabaev}, and obtained exact agreement for the
$2p_{1/2}$ and $2p_{3/2}$ states but a slightly different
result (0.001 eV) for the $2s_{1/2}$ state, which we attribute
to different treatment of the nucleus.

The next contributions to screening are the ``perturbed orbital'' (PO)
terms shown in Figs.~\ref{fig1}a and \ref{fig1}b. They are given by
\begin{equation}
    E^{\rm VP}_{\rm PO} = (U_V)_{v \tilde{v}} + (U_V)_{\tilde{v} v} +
  \sum_a \Big[ (U_V)_{a \tilde{a}} + (U_V)_{\tilde{a} a} \Big],
\label{sevp}
\end{equation}
where $U_V$ is the Uehling potential defined in Eq. (19),
\begin{eqnarray}
  \psi_{\tilde v}(\vec y) & \equiv & \alpha \sum_{m \neq v,a}
  \int {d^3 z\, d^3 w \over |\vec z - \vec w|} \,
  {\psi_m(\vec y) \over \epsilon_v - \epsilon_m} \,
  \Big[ \bar{\psi}_m(\vec z) \gamma_{\mu} \psi_v(\vec z) \,
        \bar{\psi}_a(\vec w) \gamma^{\mu} \psi_a(\vec w)
 \nonumber \\ &&
  -\,e^{i \delta E_{va} |\vec z - \vec w|} \,
        \bar{\psi}_m(\vec z) \gamma_{\mu} \psi_a(\vec z) \,
        \bar{\psi}_a(\vec w) \gamma^{\mu} \psi_v(\vec w) \Big]
 \nonumber \\ &&
  - \sum_{m \neq v} \int\! d^3 z\,
  {\psi_m(\vec y) \over \epsilon_v - \epsilon_m } \,
   \psi^\dagger_m(\vec z) U(z) \psi_v(\vec z).
\end{eqnarray}
is a valence orbital perturbed either by the exchange of a photon
with the core electrons or else by the counter potential $U(z)$,
and
\begin{eqnarray}
  \psi_{\tilde a}(\vec y)  & \equiv & \alpha \sum_{m \neq a}
  \int {d^{\,3}z\, d^{\,3}w \over |\vec z - \vec w|} \,
  {\psi_m(\vec y) \over \epsilon_a - \epsilon_m} \,
  \Big[ \bar{\psi}_m(\vec z) \gamma_{\mu} \psi_a(\vec z) \,
        \bar{\psi}_v(\vec w) \gamma^{\mu} \psi_v(\vec w)
 \nonumber \\ &&
  -\,e^{i \delta E_{va} |\vec z - \vec w|} \,
        \bar{\psi}_m(\vec z) \gamma_{\mu} \psi_v(\vec z) \,
        \bar{\psi}_v(\vec w) \gamma^{\mu} \psi_a(\vec w) \Big]
\end{eqnarray}
is a core orbital perturbed by the exchange of a photon with
the valence electron.

Finally, the ``derivative'' terms, which arise from the energy dependence
of one-photon exchange, are given by
\begin{equation}
  E^{\rm VP}_{\rm der} = -\sum_a (U_{vv} - U_{aa})
  g'_{avva}(\delta E_{av})\,.
\end{equation}
We have precise agreement with the Coulomb potential results
for lithiumlike uranium presented in Ref.~\cite{Shabaev},
where this contribution is denoted by $E_b$.

In Tables \ref{tab5} -- \ref{tab7}, we present vacuum polarization
results for the three alkalilike isoelectronic sequences mentioned
above. We choose to use atomic units in this case, and results of
the screening calculations just described are presented, along with
those of the Uehling and Wichmann-Kroll terms. It is interesting
to note that while the higher-order screening corrections are
typically smaller in size than the lowest-order Uehling energies,
such is not the case for the $np_{3/2}$ states, where the screening
corrections are consistently larger than, or at least comparable
to, the corresponding Uehling energies.

\section{Conclusions}

The calculations presented here are intended to play a role in precision
spectroscopy of highly charged ions, so it is important to be quantitative
about the accuracy. We choose to do this in the context of ions of uranium,
which are of interest because these ions have the highest nuclear charge that
has precision spectroscopy available, and relativistic and QED effects are
highly enhanced at high $Z$. The earliest precise measurement, on lithiumlike
uranium \cite{Gould}, had a precision of 0.1 eV. This experiment measured
the $2s_{1/2}-2p_{1/2}$ transition, and subsequently the $2s_{1/2}-2p_{3/2}$
transition was measured with an accuracy of 0.27 eV in Ref.~\cite{Beier1}.
More recently, the $3s - 3p_{3/2}$ transition in sodiumlike uranium has been
measured \cite{na92}, and the experimental precision has reached 0.02 eV.
Finally, the $4s - 4p_{3/2}$ transition in copperlike uranium has been
measured to the remarkable precision of 0.0019 eV \cite{cu92}.

These high precision measurements pose a considerable challenge to the
theory of many-electron ions. While correlation calculations based on
Dirac-Fock or model potentials converge quite rapidly at this high $Z$
and are presently sufficiently accurate, the proper incorporation of
negative-energy state and retardation effects is best done in a
field-theoretic context, with the relativistic many-body treatment
of correlation replaced with the evaluation of Feynman diagrams involving
the exchange of two photons between electrons. This procedure has been
carried out for lithiumlike ions \cite{LLNL2,Shabaev}. After this has been
done, one needs to carry out radiative correction calculations in the
many-electron environment, which requires both precise one-loop calculations
and associated screening corrections. One purpose of the present paper has
been to carry out the vacuum polarization part of this program with accuracy
at least at the level of the experimental uncertainties, and we now discuss
to what extent we have succeeded in this attempt.

We begin by discussing lithiumlike uranium, where the overall contribution
of vacuum polarization to the $2s_{1/2}$ energy is -14.742 eV. While the
numerical determination of the Uehling contribution to this of -15.831 eV
is extremely accurate, care must be taken with nuclear size dependence.
Varying the $c$ parameter used here, 7.13753 fm, by one percent leads
to changes of 0.008 eV, and because the uncertainty quoted by
Zumbro {\it et al.}~\cite{Zumbro92} for the $c$ parameter is
0.0012 fm (0.02 \%), this part of the calculation needs no improvement.
If experiment continues to improve, however, a more detailed study
of the effect of different distributions of charge and higher
multipoles may be necessary.

Turning to the Wichmann-Kroll calculation, which contributes 0.789 eV,
the main uncertainty here is the cutting off of the partial wave series
at $\kappa = 5$. For the case of the $n=2$ states of lithiumlike uranium,
we extended the sum to $\kappa = 10$, and found a change of only 0.001 eV.
In view of the rapid convergence of the partial wave series for this term,
stopping the calculation at $\kappa = 5$ should be quite adequate.

Finally, a more delicate issue involves screening. Here we made
one approximation that can be important, namely, treating the vacuum
polarization contribution to exchange photons in Uehling approximation.
Ref.~\cite{Shabaev} explicitly calculated the residual Wichmann-Kroll
term using a Coulomb potential, and found a 0.0005 eV contribution to
the $2s$ energy, so our approximation is valid. However, higher-order
screening graphs have not been treated here, and we therefore estimate
the size of their contributions by comparing the results of different
model potentials. We use a core-Hartree potential, defined with an
effective charge
\begin{equation}
  Z_{\rm eff}(r) = Z_{\rm nuc}(r) - r \!\int\! dr' {1 \over r_>} \rho_c(r'),
\end{equation}
where
$\rho_c(r) = \sum_a\,(2j_a + 1) \big[ g_a^2(r) + f_a^2(r) \big]$ is
the total charge density of the core electrons. In Table \ref{tab8},
we compare Kohn-Sham (KS) and core-Hartree (CH) results for all three
states. Considering the $2s$ state, we see that while a 0.096 eV change
is present in the Uehling potential, inclusion of screening reduces
this to a 0.016 eV difference, far smaller than the experimental error.
The $2p$ states are seen to be under even better control. We consider
that our results for vacuum polarization are accurate to under
0.02 eV for lithiumlike uranium.

While the experimental accuracy of sodiumlike uranium is higher than
lithiumlike uranium, the higher principal quantum number reduces the
size of vacuum polarization, so that it need be known with less precision.
Arguments similar to those given for lithiumlike uranium can be used
to show that the Uehling and Wichmann-Kroll terms are sufficiently
well determined. A comparison of the KS and CH results is given again
in Table \ref{tab8}. It can be seen that the $3s$ Uehling energy changes
by 0.054 eV, but inclusion of screening reduces this to 0.001 eV.
We estimate the accuracy of our result here to be 0.003 eV, which is
due mainly to the potential dependence of the Wichmann-Kroll term and
is again well under the experimental error.

Finally, copperlike uranium presents the greatest challenge to theory,
both because of the complexity of the ion and the very high experimental
precision that has been reached. We again compare KS and CH results in
Table \ref{tab8}, and see that $4s$ Uehling energy changes by 0.031 eV,
an order of magnitude larger than the experimental error. Inclusion of
first-order screening reduces this to 0.002 eV, same as the experimental
error. We note that the $4p_{1/2}$ energy shifts by about the same amount,
so the potential dependence of the $4s_{1/2} - 4p_{1/2}$ transition
energy is much less than the experimental error, but the shift of the
$4p_{3/2}$ level is smaller, though it still reduces the potential
dependence of the $4s_{1/2} - 4p_{3/2}$ transition energy. The shift
of the Wichmann-Kroll terms is seen to be significant on the scale of
the experimental error, but cancels when transitions are considered.
Were we to include screening corrections to the Wichmann-Kroll term,
we expect the results of the two potentials would give closer answers
for the individual states. Thus for copperlike uranium, the vacuum
polarization results presented here should be regarded as having an
accuracy comparable to experiment for transitions, but further work
is needed for ionization energies.

While the vacuum polarization calculations presented here are adequate
at present, further reduction of experimental error will require a more
accurate treatment of screening. The treatment given here is equivalent to
first-order many body perturbation theory (MBPT), though it is done in a
field theoretic manner. Further accuracy should be obtainable by inclusion
of second-order MBPT, which is a straightforward though lengthy procedure.

The next step is to treat the self-energy diagram including screening
corrections. This has been carried out for lithiumlike ions
\cite{Shabaev,LLNL2}, and the theoretical interest for this isoelectronic
sequence has shifted to the proper inclusion of the two-loop Lamb shift,
which has recently been calculated for the ground state of hydrogenic
ions \cite{Shabaev3}. While considerable work remains to extend these
calculations to the sodium and copper isoelectronic sequences,
the prospect of treating these truly many-electron systems in a purely
QED manner is quite promising.

\begin{acknowledgments}
The work of J.S. was supported in part by NSF Grant No.~PHY-0097641.
The work of K.T.C. was performed under the auspices of the U.S. Department
of Energy by the University of California, Lawrence Livermore National
Laboratory under Contract No.~W-7405-Eng-48.
We acknowledge useful conversations with P.J. Mohr.
\end{acknowledgments}


\vspace{2in}

\begin{table*}[p]
\caption{\label{tab1}
Behavior of the unrenormalized vacuum polarization for the $1s$ ground
state of hydrogenlike mercury ($Z=80$). W-K is the finite Wichmann-Kroll
term given by the sum of the first three terms. Units: a.u..}
\begin{ruledtabular}
\begin{tabular}{crrrr}
  \multicolumn{1}{c}{$\kappa$}
& \multicolumn{1}{c}{$E^{\rm VP}$}
& \multicolumn{1}{c}{$E^{\rm VP}(S_0)$}
& \multicolumn{1}{c}{Uehling}
& \multicolumn{1}{c}{W-K} \\
\colrule
-1   &  -263.566739  & -33.856852  &  229.754601            \\
~1   &  -195.878973  &  33.856852  &  229.754601            \\
Sum  &  -459.445712  &   0.000000  &  459.509201 & 0.063489 \\
[4pt]
-2   &  -515.532936  & -54.346160  &  461.190811            \\
~2   &  -406.842658  &  54.346160  &  461.190811            \\
Sum  &  -922.375594  &   0.000000  &  922.381621 & 0.006027 \\
[4pt]
-3   &  -761.361202  & -68.805626  &  692.556408            \\
~3   &  -623.750313  &  68.805626  &  692.556408            \\
Sum  & -1385.111515  &   0.000000  & 1385.112815 & 0.001300 \\
[4pt]
-4   & -1003.705497  & -79.878165  &  923.827583            \\
~4   &  -843.949263  &  79.878165  &  923.827583            \\
Sum  & -1847.654760  &   0.000000  & 1847.655166 & 0.000406 \\
[4pt]
-5   & -1243.815854  & -88.801356  & 1155.014595            \\
~5   & -1066.213175  &  88.801356  & 1155.014595            \\
Sum  & -2310.029029  &   0.000000  & 2310.029189 & 0.000160 \\
\end{tabular}
\end{ruledtabular}
\end{table*}

\begin{table*}[p]
\caption{\label{tab2}
Finite nuclear size Coulomb results for $ns$-state vacuum polarization
in terms of the function $F(Z\alpha)$.}
\begin{ruledtabular}
\begin{tabular}{rcrrrrr}
  \multicolumn{1}{c}{Z}
& \multicolumn{1}{c}{Contribution}
& \multicolumn{1}{c}{$1s_{1/2}$}
& \multicolumn{1}{c}{$2s_{1/2}$}
& \multicolumn{1}{c}{$3s_{1/2}$}
& \multicolumn{1}{c}{$4s_{1/2}$}
& \multicolumn{1}{c}{$5s_{1/2}$} \\
\colrule
30  & Uehling & -0.2386 & -0.2468 & -0.2470 & -0.2466 & -0.2462 \\ [-6pt]
    & W-K     &  0.0020 &  0.0020 &  0.0020 &  0.0021 &  0.0021 \\ [-6pt]
    & Sum     & -0.2366 & -0.2448 & -0.2450 & -0.2445 & -0.2441 \\
  [4pt]
40  & Uehling & -0.2418 & -0.2569 & -0.2572 & -0.2565 & -0.2557 \\ [-6pt]
    & W-K     &  0.0033 &  0.0035 &  0.0035 &  0.0035 &  0.0035 \\ [-6pt]
    & Sum     & -0.2385 & -0.2534 & -0.2537 & -0.2530 & -0.2522 \\
  [4pt]
50  & Uehling & -0.2507 & -0.2757 & -0.2762 & -0.2749 & -0.2736 \\ [-6pt]
    & W-K     &  0.0051 &  0.0054 &  0.0054 &  0.0054 &  0.0053 \\ [-6pt]
    & Sum     & -0.2456 & -0.2703 & -0.2708 & -0.2695 & -0.2683 \\
  [4pt]
60  & Uehling & -0.2661 & -0.3055 & -0.3061 & -0.3039 & -0.3018 \\ [-6pt]
    & W-K     &  0.0073 &  0.0081 &  0.0081 &  0.0080 &  0.0080 \\ [-6pt]
    & Sum     & -0.2588 & -0.2974 & -0.2980 & -0.2959 & -0.2938 \\
  [4pt]
70  & Uehling & -0.2896 & -0.3503 & -0.3511 & -0.3475 & -0.3440 \\ [-6pt]
    & W-K     &  0.0102 &  0.0118 &  0.0118 &  0.0116 &  0.0115 \\ [-6pt]
    & Sum     & -0.2794 & -0.3385 & -0.3393 & -0.3359 & -0.3325 \\
  [4pt]
80  & Uehling & -0.3244 & -0.4176 & -0.4184 & -0.4122 & -0.4066 \\ [-6pt]
    & W-K     &  0.0141 &  0.0172 &  0.0171 &  0.0168 &  0.0165 \\ [-6pt]
    & Sum     & -0.3103 & -0.4004 & -0.4013 & -0.3954 & -0.3901 \\
  [4pt]
82  & Uehling & -0.3332 & -0.4348 & -0.4355 & -0.4286 & -0.4224 \\ [-6pt]
    & W-K     &  0.0150 &  0.0185 &  0.0184 &  0.0181 &  0.0178 \\ [-6pt]
    & Sum     & -0.3182 & -0.4163 & -0.4171 & -0.4105 & -0.4046 \\
  [4pt]
90  & Uehling & -0.3753 & -0.5198 & -0.5199 & -0.5093 & -0.4999 \\ [-6pt]
    & W-K     &  0.0194 &  0.0252 &  0.0249 &  0.0243 &  0.0238 \\ [-6pt]
    & Sum     & -0.3559 & -0.4946 & -0.4950 & -0.4850 & -0.4761 \\
  [4pt]
92  & Uehling & -0.3882 & -0.5462 & -0.5461 & -0.5342 & -0.5238 \\ [-6pt]
    & W-K     &  0.0206 &  0.0272 &  0.0269 &  0.0262 &  0.0257 \\ [-6pt]
    & Sum     & -0.3676 & -0.5190 & -0.5192 & -0.5080 & -0.4981 \\
  [4pt]
100 & Uehling & -0.4524 & -0.6821 & -0.6800 & -0.6609 & -0.6445 \\ [-6pt]
    & W-K     &  0.0270 &  0.0378 &  0.0372 &  0.0360 &  0.0350 \\ [-6pt]
    & Sum     & -0.4254 & -0.6443 & -0.6428 & -0.6249 & -0.6095 \\
\end{tabular}
\end{ruledtabular}
\end{table*}

\begin{table*}[p]
\caption{\label{tab3}
Finite nuclear size Coulomb results for $np_{1/2}$-state vacuum polarization
in terms of the function $F(Z\alpha)$.}
\begin{ruledtabular}
\begin{tabular}{rcrrrr}
  \multicolumn{1}{c}{Z}
& \multicolumn{1}{c}{Contribution}
& \multicolumn{1}{c}{$2p_{1/2}$}
& \multicolumn{1}{c}{$3p_{1/2}$}
& \multicolumn{1}{c}{$4p_{1/2}$}
& \multicolumn{1}{c}{$5p_{1/2}$} \\
\colrule
30  & Uehling & -0.0033 & -0.0039 & -0.0041 & -0.0042 \\ [-6pt]
    & W-K     &  0.0001 &  0.0001 &  0.0001 &  0.0001 \\ [-6pt]
    & Sum     & -0.0032 & -0.0038 & -0.0040 & -0.0041 \\
  [4pt]
40  & Uehling & -0.0063 & -0.0075 & -0.0079 & -0.0080 \\ [-6pt]
    & W-K     &  0.0001 &  0.0002 &  0.0002 &  0.0002 \\ [-6pt]
    & Sum     & -0.0062 & -0.0073 & -0.0077 & -0.0078 \\
  [4pt]
50  & Uehling & -0.0111 & -0.0132 & -0.0138 & -0.0141 \\ [-6pt]
    & W-K     &  0.0003 &  0.0004 &  0.0004 &  0.0004 \\ [-6pt]
    & Sum     & -0.0108 & -0.0128 & -0.0134 & -0.0137 \\
  [4pt]
60  & Uehling & -0.0187 & -0.0221 & -0.0231 & -0.0235 \\ [-6pt]
    & W-K     &  0.0007 &  0.0009 &  0.0009 &  0.0009 \\ [-6pt]
    & Sum     & -0.0180 & -0.0212 & -0.0222 & -0.0226 \\
  [4pt]
70  & Uehling & -0.0309 & -0.0364 & -0.0379 & -0.0383 \\ [-6pt]
    & W-K     &  0.0015 &  0.0018 &  0.0018 &  0.0018 \\ [-6pt]
    & Sum     & -0.0294 & -0.0346 & -0.0361 & -0.0365 \\
  [4pt]
80  & Uehling & -0.0511 & -0.0600 & -0.0621 & -0.0625 \\ [-6pt]
    & W-K     &  0.0030 &  0.0035 &  0.0036 &  0.0036 \\ [-6pt]
    & Sum     & -0.0481 & -0.0565 & -0.0585 & -0.0589 \\
  [4pt]
82  & Uehling & -0.0567 & -0.0665 & -0.0686 & -0.0690 \\ [-6pt]
    & W-K     &  0.0034 &  0.0040 &  0.0041 &  0.0041 \\ [-6pt]
    & Sum     & -0.0533 & -0.0625 & -0.0645 & -0.0649 \\
  [4pt]
90  & Uehling & -0.0864 & -0.1008 & -0.1034 & -0.1035 \\ [-6pt]
    & W-K     &  0.0059 &  0.0068 &  0.0069 &  0.0069 \\ [-6pt]
    & Sum     & -0.0805 & -0.0940 & -0.0965 & -0.0966 \\
  [4pt]
92  & Uehling & -0.0964 & -0.1123 & -0.1149 & -0.1149 \\ [-6pt]
    & W-K     &  0.0068 &  0.0077 &  0.0078 &  0.0078 \\ [-6pt]
    & Sum     & -0.0896 & -0.1046 & -0.1071 & -0.1071 \\
  [4pt]
100 & Uehling & -0.1524 & -0.1759 & -0.1785 & -0.1772 \\ [-6pt]
    & W-K     &  0.0118 &  0.0132 &  0.0133 &  0.0132 \\ [-6pt]
    & Sum     & -0.1406 & -0.1627 & -0.1652 & -0.1640 \\
\end{tabular}
\end{ruledtabular}
\end{table*}

\begin{table*}[p]
\caption{\label{tab4}
Finite nuclear size Coulomb results for $np_{3/2}$-state vacuum polarization
in terms of the function $F(Z\alpha)$.}
\begin{ruledtabular}
\begin{tabular}{rcrrrr}
  \multicolumn{1}{c}{Z}
& \multicolumn{1}{c}{Contribution}
& \multicolumn{1}{c}{$2p_{3/2}$}
& \multicolumn{1}{c}{$3p_{3/2}$}
& \multicolumn{1}{c}{$4p_{3/2}$}
& \multicolumn{1}{c}{$5p_{3/2}$} \\
\colrule
30  & Uehling & -0.0005  & -0.0006  & -0.0007 & -0.0007 \\ [-6pt]
    & W-K     &  0.0000  &  0.0001  &  0.0001 &  0.0001 \\ [-6pt]
    & Sum     & -0.0005  & -0.0005  & -0.0006 & -0.0006 \\
  [4pt]
40  & Uehling & -0.0009  & -0.0011  & -0.0012 & -0.0012 \\ [-6pt]
    & W-K     &  0.0000  &  0.0001  &  0.0001 &  0.0001 \\ [-6pt]
    & Sum     & -0.0009  & -0.0010  & -0.0011 & -0.0011 \\
  [4pt]
50  & Uehling & -0.0013  & -0.0016  & -0.0017 & -0.0018 \\ [-6pt]
    & W-K     &  0.0001  &  0.0001  &  0.0001 &  0.0001 \\ [-6pt]
    & Sum     & -0.0012  & -0.0015  & -0.0016 & -0.0017 \\
  [4pt]
60  & Uehling & -0.0019  & -0.0023  & -0.0025 & -0.0025 \\ [-6pt]
    & W-K     &  0.0002  &  0.0002  &  0.0002 &  0.0002 \\ [-6pt]
    & Sum     & -0.0017  & -0.0021  & -0.0023 & -0.0023 \\
  [4pt]
70  & Uehling & -0.0025  & -0.0031  & -0.0033 & -0.0034 \\ [-6pt]
    & W-K     &  0.0003  &  0.0003  &  0.0004 &  0.0004 \\ [-6pt]
    & Sum     & -0.0022  & -0.0028  & -0.0029 & -0.0030 \\
  [4pt]
80  & Uehling & -0.0032  & -0.0040  & -0.0043 & -0.0044 \\ [-6pt]
    & W-K     &  0.0004  &  0.0005  &  0.0006 &  0.0006 \\ [-6pt]
    & Sum     & -0.0028  & -0.0035  & -0.0037 & -0.0038 \\
  [4pt]
82  & Uehling & -0.0034  & -0.0043  & -0.0046 & -0.0047 \\ [-6pt]
    & W-K     &  0.0005  &  0.0006  &  0.0006 &  0.0006 \\ [-6pt]
    & Sum     & -0.0029  & -0.0037  & -0.0040 & -0.0041 \\
  [4pt]
90  & Uehling & -0.0040  & -0.0052  & -0.0055 & -0.0057 \\ [-6pt]
    & W-K     &  0.0007  &  0.0008  &  0.0009 &  0.0009 \\ [-6pt]
    & Sum     & -0.0033  & -0.0044  & -0.0046 & -0.0048 \\
  [4pt]
92  & Uehling & -0.0042  & -0.0054  & -0.0058 & -0.0060 \\ [-6pt]
    & W-K     &  0.0007  &  0.0009  &  0.0010 &  0.0010 \\ [-6pt]
    & Sum     & -0.0035  & -0.0045  & -0.0048 & -0.0050 \\
  [4pt]
100 & Uehling & -0.0050  & -0.0065  & -0.0070 & -0.0072 \\ [-6pt]
    & W-K     &  0.0010  &  0.0013  &  0.0013 &  0.0014 \\ [-6pt]
    & Sum     & -0.0040  & -0.0052  & -0.0057 & -0.0058 \\
\end{tabular}
\end{ruledtabular}
\end{table*}

\begin{table*}[p]
\caption{\label{tab5}
Kohn-Sham results for vacuum polarization in the lithium isoelectronic
sequence: units a.u..}
\begin{ruledtabular}
\begin{tabular}{rcrrr}
  \multicolumn{1}{c}{Z}
& \multicolumn{1}{c}{Contribution}
& \multicolumn{1}{c}{$2s_{1/2}$}
& \multicolumn{1}{c}{$2p_{1/2}$}
& \multicolumn{1}{c}{$2p_{3/2}$} \\
\colrule
30  & Uehling   & -0.00278 & -0.00003 &  0.00000 \\ [-10pt]
    & W-K       &  0.00000 &  0.00000 &  0.00000 \\ [-10pt]
    & Screening &  0.00010 &  0.00010 &  0.00009 \\ [-10pt]
    & Sum       & -0.00268 &  0.00007 &  0.00009 \\
40  & Uehling   & -0.00939 & -0.00019 & -0.00001 \\ [-10pt]
    & W-K       &  0.00013 &  0.00001 &  0.00000 \\ [-10pt]
    & Screening &  0.00026 &  0.00025 &  0.00021 \\ [-10pt]
    & Sum       & -0.00900 &  0.00006 &  0.00021 \\
50  & Uehling   & -0.02498 & -0.00090 & -0.00008 \\ [-10pt]
    & W-K       &  0.00050 &  0.00003 &  0.00001 \\ [-10pt]
    & Screening &  0.00060 &  0.00057 &  0.00044 \\ [-10pt]
    & Sum       & -0.02388 & -0.00030 &  0.00037 \\
60  & Uehling   & -0.05795 & -0.00327 & -0.00027 \\ [-10pt]
    & W-K       &  0.00155 &  0.00013 &  0.00003 \\ [-10pt]
    & Screening &  0.00126 &  0.00119 &  0.00080 \\ [-10pt]
    & Sum       & -0.05514 & -0.00195 &  0.00055 \\
70  & Uehling   & -0.12395 & -0.01025 & -0.00074 \\ [-10pt]
    & W-K       &  0.00420 &  0.00051 &  0.00093 \\ [-10pt]
    & Screening &  0.00252 &  0.00239 &  0.00134 \\ [-10pt]
    & Sum       & -0.11722 & -0.00735 &  0.00070 \\
80  & Uehling   & -0.25327 & -0.02940 & -0.00170 \\ [-10pt]
    & W-K       &  0.01044 &  0.00174 &  0.00025 \\ [-10pt]
    & Screening &  0.00494 &  0.00482 &  0.00215 \\ [-10pt]
    & Sum       & -0.23790 & -0.02284 &  0.00071 \\
83  & Uehling   & -0.31231 & -0.03991 & -0.00213 \\ [-10pt]
    & W-K       &  0.01356 &  0.00249 &  0.00034 \\ [-10pt]
    & Screening &  0.00603 &  0.00597 &  0.00246 \\ [-10pt]
    & Sum       & -0.29271 & -0.03145 &  0.00067 \\
90  & Uehling   & -0.50673 & -0.08045 & -0.00352 \\ [-10pt]
    & W-K       &  0.02455 &  0.00555 &  0.00063 \\ [-10pt]
    & Screening &  0.00962 &  0.00996 &  0.00333 \\ [-10pt]
    & Sum       & -0.47256 & -0.06495 &  0.00044 \\
92  & Uehling   & -0.58176 & -0.09815 & -0.00404 \\ [-10pt]
    & W-K       &  0.02900 &  0.00695 &  0.00075 \\ [-10pt]
    & Screening &  0.01101 &  0.01157 &  0.00362 \\ [-10pt]
    & Sum       & -0.54175 & -0.07963 &  0.00033 \\
100 & Uehling   & -1.01615 & -0.21776 & -0.00677 \\ [-10pt]
    & W-K       &  0.05627 &  0.01692 &  0.00143 \\ [-10pt]
    & Screening &  0.01911 &  0.02165 &  0.00500 \\ [-10pt]
    & Sum       & -0.94078 & -0.17919 & -0.00034 \\
\end{tabular}
\end{ruledtabular}
\end{table*}

\begin{table*}[p]
\caption{\label{tab6}
Kohn-Sham results for vacuum polarization in the sodium isoelectronic
sequence: units a.u..}
\begin{ruledtabular}
\begin{tabular}{rcrrr}
  \multicolumn{1}{c}{Z}
& \multicolumn{1}{c}{Contribution}
& \multicolumn{1}{c}{$3s_{1/2}$}
& \multicolumn{1}{c}{$3p_{1/2}$}
& \multicolumn{1}{c}{$3p_{3/2}$} \\
\colrule
30  & Uehling   & -0.00054 & -0.00000 &  0.00000 \\ [-10pt]
    & W-K       &  0.00000 &  0.00000 &  0.00000 \\ [-10pt]
    & Screening &  0.00001 &  0.00004 &  0.00004 \\ [-10pt]
    & Sum       & -0.00053 &  0.00004 &  0.00004 \\
40  & Uehling   & -0.00206 & -0.00004 &  0.00000 \\ [-10pt]
    & W-K       &  0.00003 &  0.00000 &  0.00000 \\ [-10pt]
    & Screening &  0.00014 &  0.00013 &  0.00012 \\ [-10pt]
    & Sum       & -0.00190 &  0.00009 &  0.00012 \\
50  & Uehling   & -0.00586 & -0.00022 & -0.00001 \\ [-10pt]
    & W-K       &  0.00011 &  0.00001 &  0.00000 \\ [-10pt]
    & Screening &  0.00032 &  0.00030 &  0.00026 \\ [-10pt]
    & Sum       & -0.00542 &  0.00008 &  0.00025 \\
60  & Uehling   & -0.01418 & -0.00087 & -0.00006 \\ [-10pt]
    & W-K       &  0.00038 &  0.00004 &  0.00001 \\ [-10pt]
    & Screening &  0.00068 &  0.00064 &  0.00052 \\ [-10pt]
    & Sum       & -0.01312 & -0.00020 &  0.00045 \\
70  & Uehling   & -0.03120 & -0.00284 & -0.00019 \\ [-10pt]
    & W-K       &  0.00105 &  0.00014 &  0.00003 \\ [-10pt]
    & Screening &  0.00137 &  0.00131 &  0.00096 \\ [-10pt]
    & Sum       & -0.02880 & -0.00142 &  0.00076 \\
80  & Uehling   & -0.06505 & -0.00839 & -0.00047 \\ [-10pt]
    & W-K       &  0.00266 &  0.00049 &  0.00008 \\ [-10pt]
    & Screening &  0.00268 &  0.00264 &  0.00172 \\ [-10pt]
    & Sum       & -0.05974 & -0.00532 &  0.00124 \\
83  & Uehling   & -0.08060 & -0.01146 & -0.00060 \\ [-10pt]
    & W-K       &  0.00348 &  0.00071 &  0.00010 \\ [-10pt]
    & Screening &  0.00328 &  0.00326 &  0.00204 \\ [-10pt]
    & Sum       & -0.07389 & -0.00757 &  0.00143 \\
90  & Uehling   & -0.13199 & -0.02338 & -0.00103 \\ [-10pt]
    & W-K       &  0.00634 &  0.00158 &  0.00019 \\ [-10pt]
    & Screening &  0.00525 &  0.00536 &  0.00302 \\ [-10pt]
    & Sum       & -0.12048 & -0.01654 &  0.00199 \\
92  & Uehling   & -0.15186 & -0.02858 & -0.00120 \\ [-10pt]
    & W-K       &  0.00750 &  0.00199 &  0.00023 \\ [-10pt]
    & Screening &  0.00602 &  0.00620 &  0.00337 \\ [-10pt]
    & Sum       & -0.13844 & -0.02053 &  0.00220 \\
100 & Uehling   & -0.26701 & -0.06381 & -0.00209 \\ [-10pt]
    & W-K       &  0.01463 &  0.00484 &  0.00045 \\ [-10pt]
    & Screening &  0.01051 &  0.01132 &  0.00530 \\ [-10pt]
    & Sum       & -0.24205 & -0.04785 &  0.00329 \\
\end{tabular}
\end{ruledtabular}
\end{table*}

\begin{table*}[p]
\caption{\label{tab7}
Kohn-Sham results for vacuum polarization in the copper isoelectronic
sequence:  units a.u..}
\begin{ruledtabular}
\begin{tabular}{rcrrr}
  \multicolumn{1}{c}{Z}
& \multicolumn{1}{c}{Contribution}
& \multicolumn{1}{c}{$4s_{1/2}$}
& \multicolumn{1}{c}{$4p_{1/2}$}
& \multicolumn{1}{c}{$4p_{3/2}$} \\
\colrule
60  & Uehling   & -0.00351 & -0.00021 & -0.00001 \\ [-6pt]
    & W-K       &  0.00009 &  0.00000 &  0.00000 \\ [-6pt]
    & Screening &  0.00031 &  0.00025 &  0.00021 \\ [-6pt]
    & Sum       & -0.00311 &  0.00004 &  0.00019 \\
  [4pt]
70  & Uehling   & -0.00851 & -0.00076 & -0.00005 \\ [-6pt]
    & W-K       &  0.00029 &  0.00004 &  0.00001 \\ [-6pt]
    & Screening &  0.00064 &  0.00054 &  0.00041 \\ [-6pt]
    & Sum       & -0.00758 & -0.00019 &  0.00037 \\
  [4pt]
80  & Uehling   & -0.01890 & -0.00243 & -0.00013 \\ [-6pt]
    & W-K       &  0.00077 &  0.00014 &  0.00002 \\ [-6pt]
    & Screening &  0.00128 &  0.00113 &  0.00080 \\ [-6pt]
    & Sum       & -0.01686 & -0.00117 &  0.00067 \\
  [4pt]
83  & Uehling   & -0.02377 & -0.00337 & -0.00017 \\ [-6pt]
    & W-K       &  0.00103 &  0.00021 &  0.00003 \\ [-6pt]
    & Screening &  0.00157 &  0.00141 &  0.00096 \\ [-6pt]
    & Sum       & -0.02119 & -0.00178 &  0.00079 \\
  [4pt]
90  & Uehling   & -0.04006 & -0.00711 & -0.00031 \\ [-6pt]
    & W-K       &  0.00192 &  0.00048 &  0.00006 \\ [-6pt]
    & Screening &  0.00251 &  0.00234 &  0.00148 \\ [-6pt]
    & Sum       & -0.03565 & -0.00433 &  0.00117 \\
  [4pt]
92  & Uehling   & -0.04641 & -0.00877 & -0.00037 \\ [-6pt]
    & W-K       &  0.00229 &  0.00061 &  0.00007 \\ [-6pt]
    & Screening &  0.00287 &  0.00271 &  0.00167 \\ [-6pt]
    & Sum       & -0.04128 & -0.00549 &  0.00131 \\
  [4pt]
100 & Uehling   & -0.08348 & -0.02009 & -0.00067 \\ [-6pt]
    & W-K       &  0.00457 &  0.00152 &  0.00015 \\ [-6pt]
    & Screening &  0.00495 &  0.00495 &  0.00272 \\ [-6pt]
    & Sum       & -0.07402 & -0.01369 &  0.00209 \\
\end{tabular}
\end{ruledtabular}
\end{table*}

\begin{table*}[p]
\caption{\label{tab8}
Comparing Kohn-Sham (KS) with core-Hartree (CH) results for
lithiumlike, sodiumlike and copperlike uranium: units eV.}
\begin{ruledtabular}
\begin{tabular}{cccrccr}
\multicolumn{1}{c}{Ion}
& \multicolumn{1}{c}{State}
& \multicolumn{1}{c}{Potential}
& \multicolumn{1}{c}{Uehling}
& \multicolumn{1}{c}{W-K~~}
& \multicolumn{1}{c}{Screening}
& \multicolumn{1}{c}{Sum} \\
\colrule
Li-like & $2s_{1/2}$ & KS & -15.831 & 0.789 & 0.300 & -14.742 \\
        &            & CH & -15.735 & 0.784 & 0.220 & -14.731 \\
[4pt]
        & $2p_{1/2}$ & KS &  -2.671 & 0.189 & 0.315 &  -2.167 \\
        &            & CH &  -2.614 & 0.185 & 0.259 &  -2.170 \\
[4pt]
        & $2p_{3/2}$ & KS &  -0.110 & 0.020 & 0.099 &   0.009 \\
        &            & CH &  -0.109 & 0.020 & 0.097 &   0.008 \\
[8pt]
Na-like & $3s_{1/2}$ & KS &  -4.132 & 0.204 & 0.164 &  -3.764 \\
        &            & CH &  -4.078 & 0.201 & 0.109 &  -3.768 \\
[4pt]
        & $3p_{1/2}$ & KS &  -0.778 & 0.054 & 0.169 &  -0.555 \\
        &            & CH &  -0.755 & 0.052 & 0.141 &  -0.562 \\
[4pt]
        & $3p_{3/2}$ & KS &  -0.033 & 0.006 & 0.092 &   0.065 \\
        &            & CH &  -0.025 & 0.006 & 0.085 &   0.066 \\
[8pt]
Cu-like & $4s_{1/2}$ & KS &  -1.263 & 0.063 & 0.077 &  -1.123 \\
        &            & CH &  -1.232 & 0.061 & 0.044 &  -1.127 \\
[4pt]
        & $4p_{1/2}$ & KS &  -0.239 & 0.017 & 0.073 &  -0.149 \\
        &            & CH &  -0.229 & 0.017 & 0.061 &  -0.151 \\
[4pt]
        & $4p_{3/2}$ & KS &  -0.010 & 0.002 & 0.044 &   0.036 \\
        &            & CH &  -0.010 & 0.002 & 0.043 &   0.035 \\
\end{tabular}
\end{ruledtabular}
\end{table*}

\end{document}